\documentclass[twocolumn,showkeys,superscriptaddress,preprintnumbers,amsmath,amssymb,showpacs,prl]{revtex4-1}
\usepackage{graphicx}
\usepackage{dcolumn}
\usepackage{bm}
\usepackage{amsmath}
\usepackage{color} 
\begin{document}

\title{Crackling to periodic dynamics in sheared granular media} 

\author{Aghil Abed Zadeh} \email{aghil.abed.zadeh@duke.edu}
\affiliation{Department of Physics \& Center for Non-linear and Complex Systems, Duke University, Durham, NC, USA}
\author{Jonathan Bar{\'e}s} \email{jb@jonathan-bares.eu}
\affiliation{Department of Physics \& Center for Non-linear and Complex Systems, Duke University, Durham, NC, USA}
\affiliation{Laboratoire de M\'{e}canique et G\'{e}nie Civil, Universit\'{e} de Montpellier, CNRS, Montpellier, France}
\author{Robert P. Behringer}
\affiliation{Department of Physics \& Center for Non-linear and Complex Systems, Duke University, Durham, NC, USA}


\begin{abstract}
We study the local and global dynamics of sheared granular materials in a model experiment. The system crackles, with intermittent slip avalanches, or exhibits periodic motion, depending on the shear rate or loading stiffness. The global force on the intruder during shearing captures the transition from the crackling to the periodic regime. We deduce a novel dynamic phase diagram as a function of the shear rate and the system's stiffness and associated scaling laws. Using photo-elasticimetry, we also capture the grain-scale stress evolution, and investigate the microscopic behavior in the different regimes.  
\end{abstract}

\date{\today}

\keywords{crackling dynamics, periodic dynamics, granular, avalanches, stick-slip}

\pacs{45.70.-n 91.30.pa 62.20.F 45.70.Ht} 

\maketitle


Sheared amorphous materials yield and flow, when sufficiently loaded \cite{regev2015reversibility,lin2015criticality,maloney2006amorphous}. The flow can be spatially heterogeneous and erratic in time. This intermittent behavior has been observed in phenomena as diverse as seismicity \cite{Bak02_prl,davidsen2013_prl,bares2018_natcom}, fracture \cite{Alava06_ap,Bonamy08_prl,bares14_prl}, damage \cite{petri1994_prl,ribeiro2015_prl}, friction \cite{brace1966stick,johnson2008effects}, plasticity \cite{zapperi1997_nat,Papanikolaou2012_nat}, magnetization \cite{Urbach95_prl,Durin05_book}, wetting \cite{ertacs1994_pre,Planet09_prl}, neural activity \cite{beggs2003_jn,bellay2015_elife} and granular avalanches \cite{miller1996stress,Hayman2011_pag,pudasaini2007_book,lebouil_prl2014,bares2017_pre}. 

In granular media, the intermittent dynamics, also called `crackling' \cite{sethna2001crackling}, is associated with the transition between jammed and unjammed states \cite{bi2011jamming,bares2017_pre} around yielding: when slowly sheared, a granular system may stick and slip, with slip sizes spanning a wide range of scales \cite{howell_prl1999,albert2001stick,petri_epjb2008,lherminier2016granular}. Meanwhile, under certain conditions, some sheared systems do not obey this crackling behavior. Instead, slips occur periodically with a narrow size distribution \cite{nasuno1998time,nasuno1997friction,kaproth2013slow,krim2011stick}. For other systems, the oscillations may be damped or the grains flow smoothly. In addition to safety and industrial challenges to control these dynamics, understanding these different behaviors is crucial in granular physics and many other associated fields, exhibiting similar dynamics.

The crackling response of granular media is of particular interest \cite{dahmen2011simple,lin2015criticality,bares2017_pre} and the effect of the system's parameters on the dynamics, their scaling laws and avalanche shapes have been investigated \cite{aharonov2004stick,liu2016driving,white2003driving,luan2004effect}. Other studies concentrated on understanding the transition between the periodic stick-slip and steady sliding regimes \cite{lacombe2000dilatancy,krim2011stick}. However, the transition from crackling to periodic dynamics have not been studied. Consequently, fundamental issues remain: Do crackling and periodic dynamics arise from fundamentally different systemic behaviors, identifiable at a microscopic scale, or do they occur as bifurcations controlled by system-scale parameters? What is the interplay between the grain-scale mechanics and the macroscopic system parameters?   

The study reported here addresses these questions. Experiments on sheared granular materials reproduce both crackling and periodic regimes. Adjusting the driving rate and the system stiffness causes the system to transition between different behaviors, and yields a dynamic phase diagram. The system observables, such as loading force are analyzed in both regimes. We find that the periodic behavior of macroscopic force is accompanied with local erratic stress dynamics.


\begin{figure}
\centering \includegraphics[width=0.60\columnwidth]{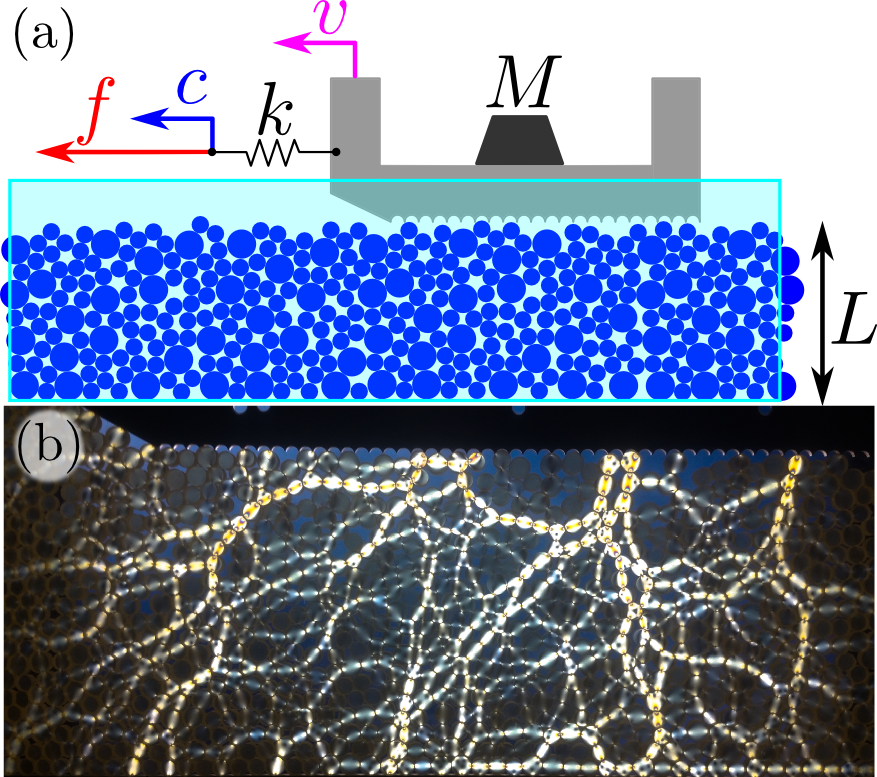}
\caption{(color online) (a) Sketch of the experiment from the side: A slider is pulled over a 2D vertical granular bed of bi-disperse photoelastic discs. The slider is pulled at a constant speed $c$ by mean of a spring of stiffness $k$. The pulling force, $f$, is measured by a force sensor, connected to the spring. The system is lit from behind by polarized light and observed from the front by a fast ($\sim 120$~fps) camera equipped with a crossed polarizer. A second camera (without polarizer) simultaneously records the particles. (b) Photoelastic response of the granular bed during loading.}
\label{fig1}
\end{figure}

\noindent \textit{Experiments} -- The experimental device provides data both at the global and local scales, similarly to \cite{krim2011stick,zadeh2017avalanches}. As shown in fig.\ref{fig1}(a) \footnote{see SM for a full picture of the experiment (Fig. S1).}, a stage pulls a 2D frictional slider of fixed length $25$~cm and variable mass $M$. The stage, which moves at constant speed $c$, pulls the slider by means of a linear spring of stiffness $k$. The slider rests on a vertical bed of fixed depth $L=9.5$~cm, and length $1.5$~m consisting of bi-disperse cylindrical photoelastic particles with diameters $0.4$~cm and $0.5$~cm (small/big ratio of $2.7$) to avoid crystallization. Unless specified, the experiments are made with a slider of mass $M=85$~g. The slider+particles system is sandwiched between two dry-lubricated glass plates. The slider bottom is toothed to enhance the friction with the grains. The force, $f$, applied to the spring, is measured by a sensor at a frequency of $1$~kHz. The system is designed to be at constant pressure, and the slider can move in either the horizontal or vertical directions. However, the granular bed is prepared flat enough that it stays mostly horizontal, while pulled. The system is lit from behind by a polarized light source. In front, a camera with a crossed polarizer images the grains and slider at a frequency of $120$~Hz (see fig.\ref{fig1}(b)). The photoelastic response of the media provides a local measure of the stress from the image intensity $I$ \cite{howell_prl1999}.

\begin{figure}
\centering \includegraphics[width=0.65\columnwidth]{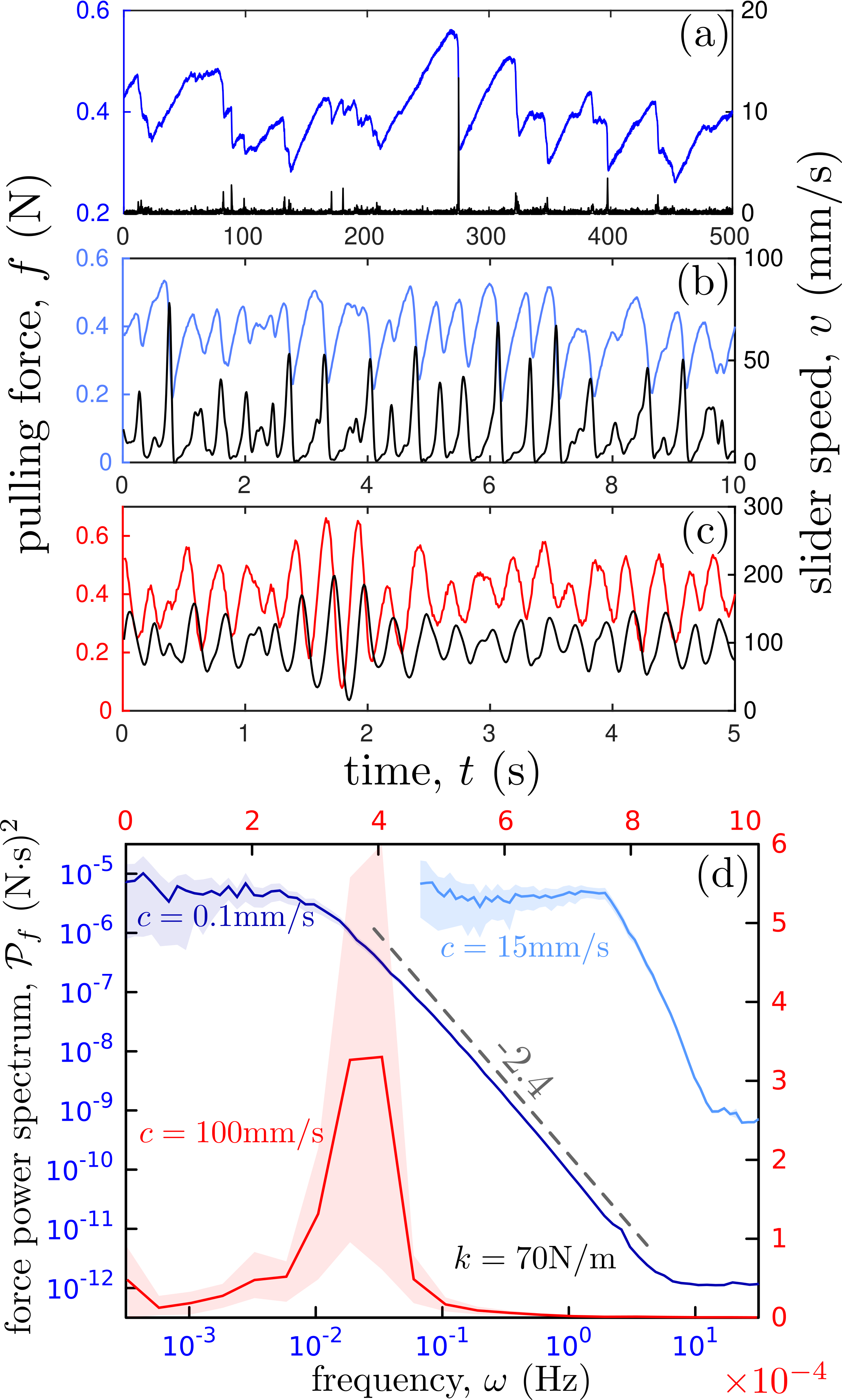}
\caption{(color online) Typical signals. Evolution of the slider speed $v(t)$ (black) and pulling force $f(t)$ (colored) for different loading speeds ($c=0.1$, $15$ and $100$~mm/s) displaying different dynamics: crackling at low speed (a), irregular at medium speed (b) and periodic at high speed (c). (d): Corresponding power spectral density of the force signal $\mathcal{P}_f$. A dashed line shows a slope corresponding with the exponent $-2.4 \pm 0.2$ fitted on the power-law obtained for $c=0.1$~mm/s. The shaded areas show the $95$\% confidence interval of the curves. The blue curves are in log-log space. Experiments are carried out with a spring of stiffness $k=70$~N/m.}
\label{fig2}
\end{figure}

\noindent \textit{Results} -- Fig.\ref{fig2} shows the evolution of the pulling force, $f(t)$ (colored), and of the slider speed $v(t)$ (black), for three typical pulling speeds. At low speed (a) $c=0.1$~mm/s, the system crackles \cite{sethna2001crackling,zadeh2017avalanches}. The slider is immobile most of the time, as it loads up elastically. It then undergoes erratic sudden jumps of high intensity. These unloading events are accompanied by slider slips and irreversible grain flows. At high speed (c) $c=100$~mm/s, the slider never stops, and exhibits smooth periodic oscillations. In between (b) $c=15$~mm/s, most of the time, the slider exhibits slow noisy displacements with erratic jumps, smoother and less intense than in the crackling case. Fig.\ref{fig2}(d) shows the power spectral density (PSD) of the force signal $\mathcal{P}_f$ for these three different cases. At low $c$, above a flat lower cut-off, $\mathcal{P}_f$ follows a power-law spanning more than two decades in $\omega$ with exponent $-2.4\pm0.2$, similar to Brownian noise \cite{baldassarri2006brownian}. At a medium $c$, $\mathcal{P}_f$ is constant for a large range of $\omega$, like white noise, and decays rapidly above a cut-off frequency ($\sim 10$~Hz). For higher $c$, $\mathcal{P}_f$ develops a peak with a characteristic frequency which depends on $c$, and differs from the constant inertial frequency of the system, $\omega_{\text{syst}}=\frac{1}{2\pi} \sqrt{\frac{k}{M}}=4.6$~Hz.

\begin{figure}
\centering \includegraphics[width=0.7 \columnwidth]{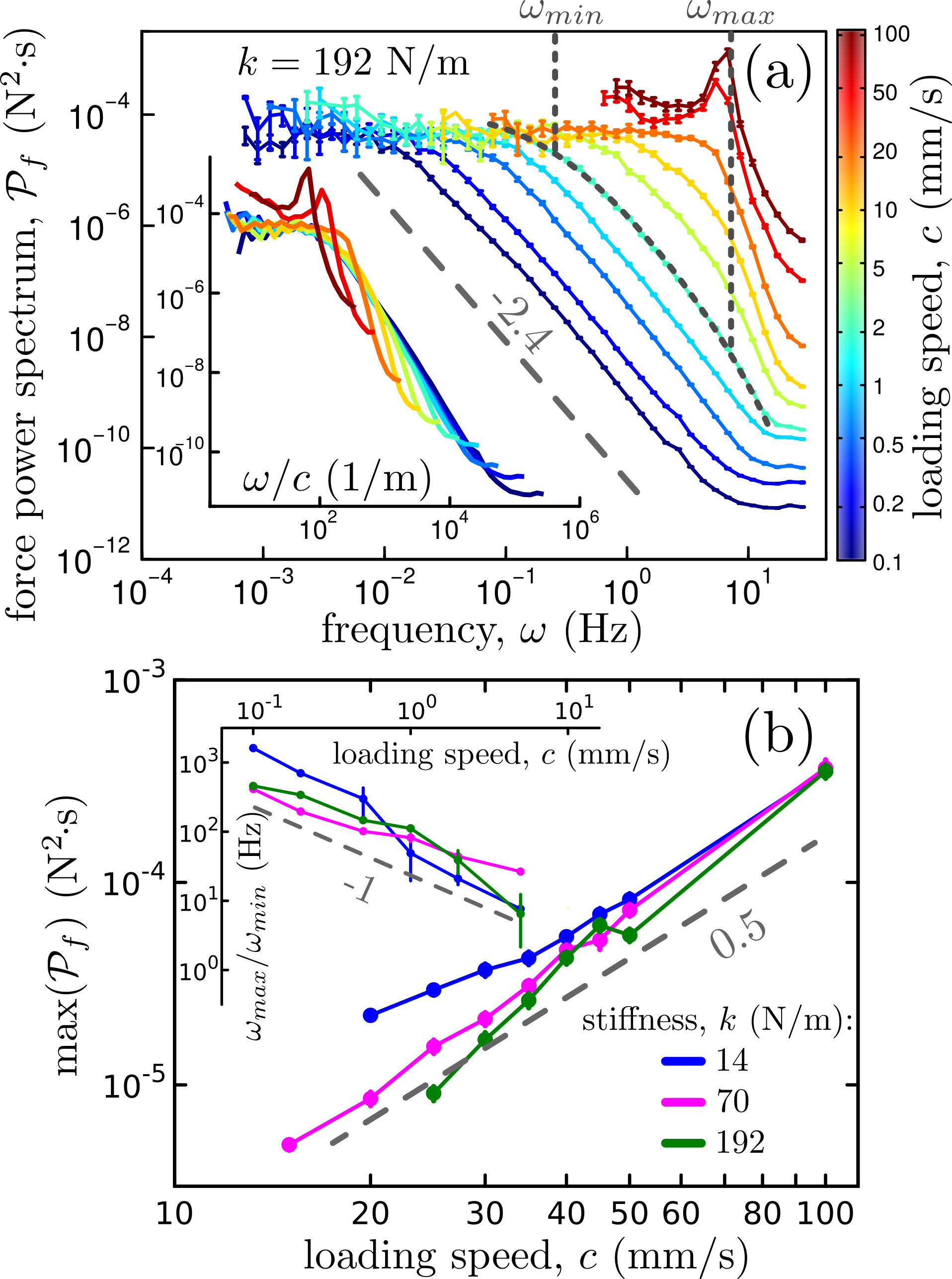}
\caption{(color online) (a): Force power spectra $\mathcal{P}_f(\omega)$ for different loading speeds, $c \in [0.1,100]$~mm/s with $k=192$~N/m. For $c=2$~mm/s, $\mathcal{P}_f$ is fitted by a gamma function displayed by a dashed line with lower ($\omega_{min}$) and upper ($\omega_{max}$) cut-off positions. A straight dashed line shows a slope corresponding with the exponent $-2.4$ fitted to the lowest $c$ curve. Inset: $\mathcal{P}_f$ with frequency scaled by $c$. (b) Maximum of $\mathcal{P}_f$ as a function of $c$, for power spectra with a bump. A dashed line with slope $0.5$ is given to guide the eye. Inset: Spreading of the power-law regime (if any) $\omega_{max}/\omega_{min}$ as a function of $c$. A dashed line with slope $-1$ is given to guide the eye. Different curves correspond to different stiffnesses, $k$.}
\label{fig3}
\end{figure}

The system can transit from the \textit{crackling} to \textit{irregular} to \textit{periodic}. Fig.\ref{fig3}(a) shows the evolution of the force PSD, $\mathcal{P}_f$, as $c$ varies, for fixed $k$ \footnote{see SM for at at other $k$ values (Fig. S2).}. Considering the crackling regime (small $c$) and ignoring the flat part of the PSD at high frequencies (corresponding with sensor limits), $\mathcal{P}_f$ is fitted to a gamma-distribution: $\mathcal{P}_f(\omega) \propto (1+\omega/\omega_{min})^{-\beta}e^{-\omega/\omega_{max}}$, where $\omega_{min}$ and $\omega_{max}$ are the lower and upper power-law cut-offs respectively, and $\beta=2.4\pm0.2$. The number of decades for which $\mathcal{P}_f$ obeys a power-law, $\log_{10}{(\omega_{max}/\omega_{min})}$, decreases as $c$ increases. This is quantified in the inset of fig.\ref{fig3}(b), showing that this number is roughly inversely proportional to $c$: $\omega_{max}/\omega_{min} \propto 1/c$. Moreover, in the inset of fig.\ref{fig3}(a), $\mathcal{P}_f$ curves, except for their upper cut-offs, collapse when $\omega$ is scaled by $c$. This implies that $\omega_{min}=\kappa c$, where $\kappa$ is a characteristic wave number of the system, and independent of $c$. For higher $c$, a bump appears close to $\omega_{min}$. This indicates the onset of oscillations at a frequency $\omega_c$. The inset of fig.\ref{fig3}(a) shows that, unlike $\omega_{min}$, $\omega_c$ does not scale with $c$, since the peaks do not collapse. Fig.\ref{fig3}(b) shows that the value of $\mathcal{P}_f$ at the peak is sub-linear in $c$, with an exponent close to $0.5$. This scaling quantifies the oscillation strength.

\begin{figure}
\centering \includegraphics[width=1\columnwidth]{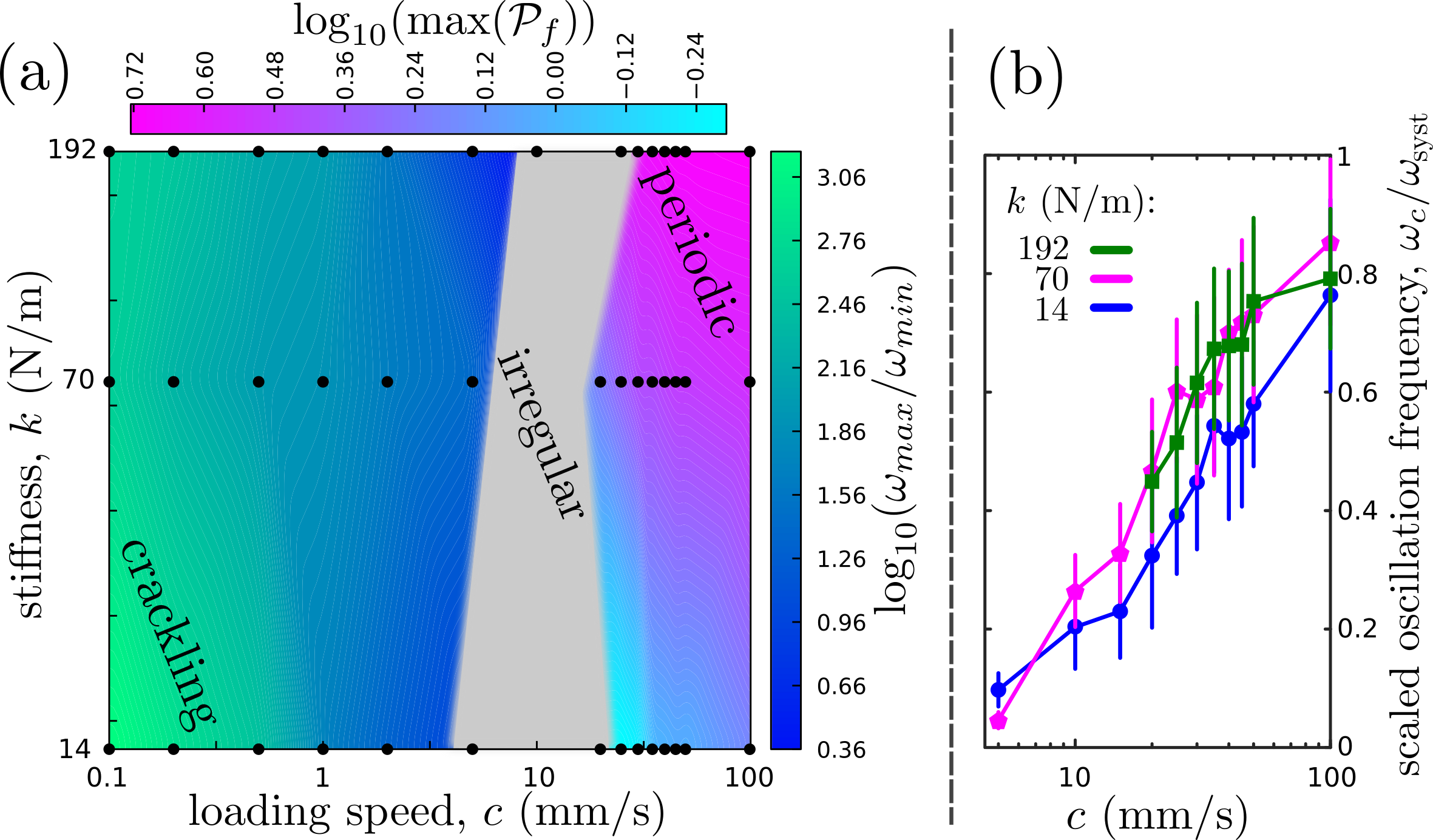}
\caption{(color online) (a) Dynamics phase diagram. The color plot of the crackling regime uses the color bar on the right. It shows the power-law range $\omega_{max}/\omega_{min}$ (left) as a function of $k$ and $c$. The color plot of the periodic regime uses the color bar on the top. It is a map of the maximum of $\mathcal{P}_f$ in the periodic regime as a function of $k$ and $c$. The irregular region, shown in gray does not have well defined periodic or crackling behavior. Black dots are experimental measurement points, the rest of the map is linearly interpolated (in the log space). (b) Evolution of PSD peak frequency, $\omega_c$ in the periodic regime, scaled by $\omega_\text{syst}$, \textit{vs.} $c$ on log-lin scales. The scaling collapses the curves for different stiffnesses.}
\label{fig4}
\end{figure}

Fig.\ref{fig4}(a) presents the phase diagram that quantify crackling and periodicity as a function of the loading speed, $c$, and the system stiffness, $k$. The number of decades ($\log_{10}{(\omega_{max}/\omega_{min})}$) is higher for lower $c$ and $k$. In this domain the crackling regime dominates. Conversely, the value of $\mathcal{P}_f$ at the peak is higher for higher $c$ and $k$. In this domain the periodic regime dominates. Between these regions lies the irregular regime, where there is neither a clearly defined power-law span nor a well defined spectral peak. The transition between these regimes is similar to an intermittent transition \cite{pomeau1980intermittent}. However, we do not detect a critical $c$ between crackling and periodic regimes. Fig.\ref{fig4}b shows how the oscillation frequency evolves with $c$ for different $k$'s. We observe a logarithmic dependence, $\omega_c \propto \log{c}$. This relation is consistent with a homoclinic bifurcation with $c$ \cite{strogatz2014nonlinear}. Furthermore, the curves for different stiffnesses collapse when $\omega_c$ is scaled by $\omega_{\text{syst}}=\frac{1}{2\pi} \sqrt{\frac{k}{M}}$ \footnote{see SM for unscaled curves (Fig. S3).}. 

\begin{figure}
\centering \includegraphics[width=0.75\columnwidth]{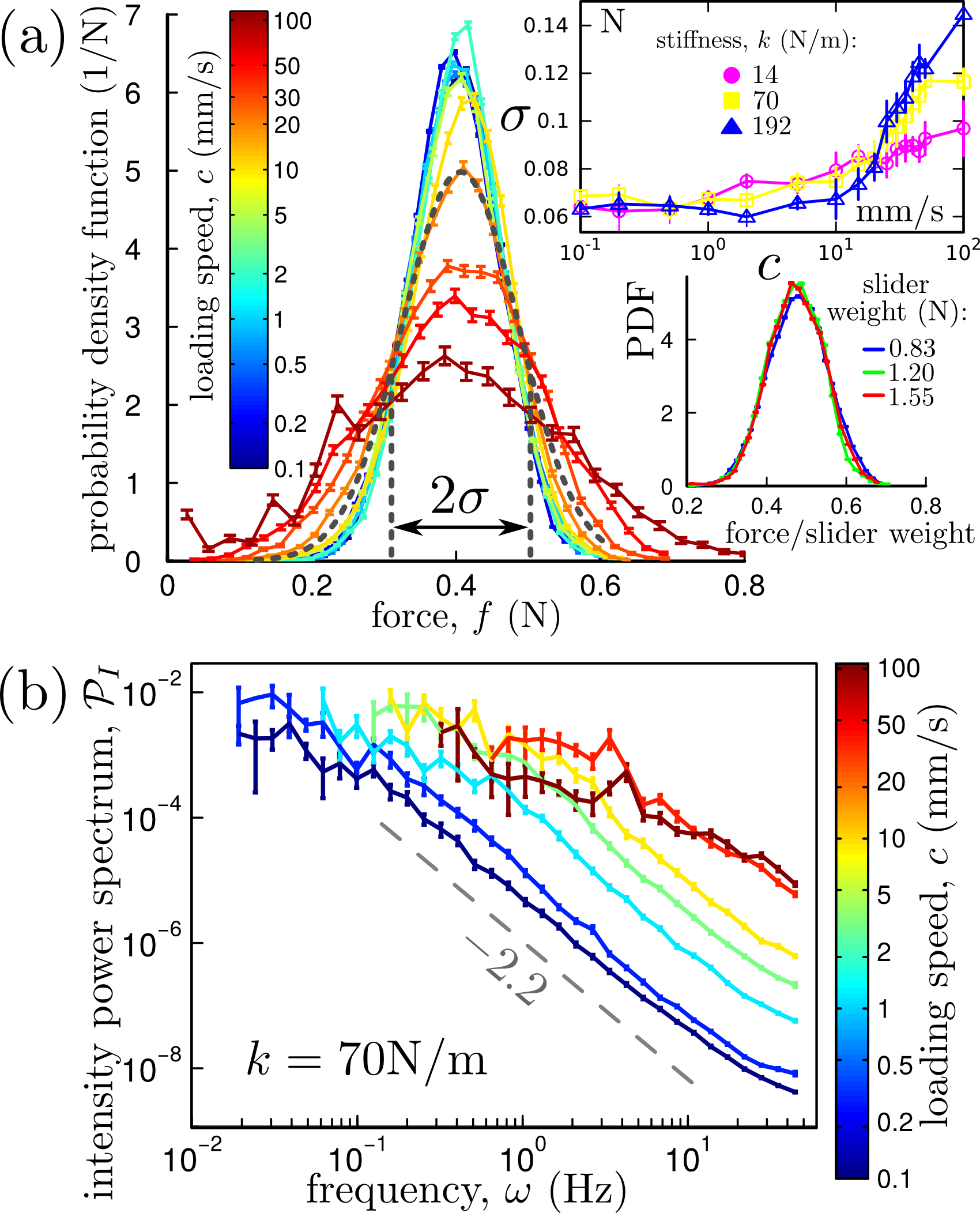}
\caption{(color online) (a) Probability density function of force $P(f)$, for different loading speeds, $c$, at fixed stiffness $k=192$~N/m. For $c=20$~mm/s, $P(f)$ is fitted by a Gaussian displayed by a dashed line. The standard deviation is $\sigma$. Inset-up: $\sigma$ \textit{vs.}\ $c$ for different stiffnesses, $k \in \{14,70,192\}$~N/m. Inset-down: PDF of the force signal scaled by the slider weight for different weights, $Mg \in \{0.83,1.2,1.55\}$~N with $c=0.5$~mm/s and $k=192$~N/m. (b) Image intensity ($I(t)$) power spectra, $\mathcal{P}_I(\omega)$, for different $c \in [0.1,100]$~mm/s with $k=70$~N/m. A straight dashed line shows a slope corresponding with the exponent $-2.2 \pm 0.2$ fitted to the lowest speed.}
\label{fig5}
\end{figure}

We also investigate the force signal probability distribution function (PDF), $P(f)$. Fig.\ref{fig5}(a) shows $P(f)$ for different $c$, and fixed $k$. Regardless of the shear rate, these PDF's are well fitted by a Gaussian function of standard deviation $\sigma(c,k)$. Their mean value is independent of $c$ in the range of speeds explored here. Similarly, as shown in the upper inset of fig.\ref{fig5}(a), the force fluctuations, quantified by $\sigma$, remain constant ($\sigma \approx 0.065$) in the crackling and irregular regimes. However, $\sigma$ starts to increase when the periodic behavior is first observed (from $c \approx 20$~mm/s), exhibiting more fluctuations.

We also explore the effect of other macroscopic parameters. Unlike previous findings \cite{liu2016driving}, changing the granular layer depth, $L$, does not change the system's behavior \footnote{see SM for $\mathcal{P}_f$ and $P(f)$ for different $L$ (Fig. S5).} to our experimental precision. The slider motion creates a narrow shear band, and particle flow is limited to the few top layers of grains. The narrowness of this shear band, where plastic granular flow occurs, is likely responsible for the fact that $L$ does not play a significant role in our experiment. However, the granular global pressure, from the slider weight, $W=Mg$, changes $P(f)$ significantly. The lower inset of fig.\ref{fig5}(a) shows these PDF's when the force is scaled by the slider mass, $f/W$. $P(f/W)$ for different $M$ collapse on a single Gaussian curve, indicating that $M$ controls both the mean and fluctuations of $f$ \footnote{see SM for non-scaled curves (Fig. S4).}. This collapse also implies a linear relation between the global pressure, $\mathcal{P}$, and the mean global shear stress, $\tau$, and its fluctuations $\sigma_{\tau}$. This provides a constant average friction coefficient, $\mu=\bar{\tau}/\mathcal{P} \approx 0.47$, for the slowly sheared medium.

We also study the relation between the global dynamics, as characterized by the $f(t)$, and the local dynamics determined by the granular stress. The local stress is computed from the polarized images intensity, $I$ \footnote{see SM for the calibration between $f(t)$ and $I(t)$ (Fig. S7).}. Fig.\ref{fig5}(b) shows the image intensity PSD, $\mathcal{P}_I$, for several $c$ at fixed $k=70$~N/m. Like $\mathcal{P}_f$ in the crackling regime, $\mathcal{P}_I$ follows a power-law with exponent $-2.2 \pm 0.2$, and the frequency interval of this power-law regime decreases as $c$ increases. However, as $c$ increases, we do not observe a clear peak in $\mathcal{P}_I$, that would indicate a periodic regime. The granular material is significantly stiffer than the spring, and the particles have lower masses. For high $c$, these differences result in interior granular fluctuations that are much faster than the global fluctuations associated with the slider and spring \footnote{see SM for a comparison between $f(t)$ and $I(t)$ signals for different loading speeds (Fig. S6).}. These granular stress fluctuations look almost the same for different loading speeds. In the periodic regime, the local stress fluctuations decouple from the flow of top layer particles, which is correlated with the global force. This decoupling occurs in the presence of an external spring, as for $k \rightarrow \infty$, we do not observe a periodic regime \footnote{see SM for $k=\infty$ PSD (Fig. S6).}.


\noindent \textit{Concluding discussion} -- The experiments reported here demonstrate crackling and periodic dynamics while shearing granular matter via a slider and a spring. The system exhibits crackling, irregular or periodic, depending on the stage speed, $c$, and the spring stiffness, $k$. The transition from crackling to periodic dynamics occurs over a region of $c-k$-space, without a sharp crossover. The other main observations of these experiments are: ($i$) The shearing force PSD provides a useful tool to indicate the dynamics of the system for a given set of system control parameters, namely loading speed, system stiffness, global pressure, and system size. ($ii$) The transition to periodic regime is similar to a homoclinic bifurcation, where $c$ is the primary control parameter. ($iii$) The shearing force PDF follows a Gaussian distribution with a standard deviation, $\sigma$, that bifurcates at the onset of periodicity; $\sigma$ is nearly independent of $c$ below the periodic regime, and grows sharply with $c$ above the transition. ($iv$) No clear periodicity is observed for the local stress dynamics, including when the slider is in the periodic regime. 

These observations are in agreement with various numerical simulations. Lacombe et al. \cite{lacombe2000dilatancy} introduced a simple friction model, similar to our experiment, with  two degrees of freedom, considering dilation. They reproduced stick-slip, inertial oscillations and sliding regimes by increasing the shear rate. However, no crackling dynamics was observed, since their model lacks stochasticity. Our phase diagram, presented in fig.\ref{fig4}, may be a detailed version of the stick-slip domain of their diagram. In an experimental study, Kaproth et al. \cite{kaproth2013slow} observed only periodic stick-slip and sliding regimes, and the driving rate only affects the slip event period. We believe they do not observe crackling dynamics because of mono-dispersity of their granular layers. In other studies, Aharonov et al. \cite{aharonov2004stick} simulated a system, very similar to ours, using the discrete element method. They also observed crackling, oscillatory and sliding regimes by changing the loading speed and the slider mass. However, the effect of the stiffness was not tested and the exact domain of the crackling regime were not investigated. Liu et al.\ \cite{liu2016driving} also studied the effect of the driving rate on avalanches. As in our experiments, they observed that increasing the driving rate decreases the power-law range of the avalanche size distribution, \textit{i.e.} the number of decades over which the avalanche size PDF obeys a power-law. Avalanches have also been observed in depinning models and a phase diagram of dynamical regimes for such a model has been demonstrated \cite{bares2013_prl} as a function of $c$, $k$ and $L$.

The findings of this letter can inform a number of open problems, ranging from avalanche dynamics for non-zero shear rate, to rheology of the sheared granular media, fracture dynamics and the depinning transition. Future work is required to investigate global and local (space-time) avalanche statistics in the crackling regime.


Acknowledgements: We would like to acknowledge Mark Robbins, Sid Nagel and Josh Socolar for their helpful conversations and funding support from NSF-DMR1206351, NASA NNX15AD38G, The William M. Keck Foundation and DARPA grant 4-34728.


\bibliographystyle{unsrt}
\bibliography{biblio.bib}

\begin{thebibliography}{10}

\bibitem{regev2015reversibility}
I.~Regev, J.~Weber, C.~Reichhardt, K.~A. Dahmen, and T.~Lookman.
\newblock Reversibility and criticality in amorphous solids.
\newblock {\em Nature communications}, 6, 2015.

\bibitem{lin2015criticality}
J.~Lin, T.~Gueudr{\'e}, A.~Rosso, and M.~Wyart.
\newblock Criticality in the approach to failure in amorphous solids.
\newblock {\em Physical review letters}, 115(16):168001, 2015.

\bibitem{maloney2006amorphous}
C.~E. Maloney and A.~Lema{\^\i}tre.
\newblock Amorphous systems in athermal, quasistatic shear.
\newblock {\em Physical Review E}, 74(1):016118, 2006.

\bibitem{Bak02_prl}
P.~Bak, K.~Christensen, L.~Danon, and T.~Scanlon.
\newblock Unified scaling law for earthquakes.
\newblock {\em Physsical Review Letter}, 88(17):178501, Apr 2002.

\bibitem{davidsen2013_prl}
J.~Davidsen and G.~Kwiatek.
\newblock Earthquake interevent time distribution for induced micro-, nano-,
  and picoseismicity.
\newblock {\em Physical review letters}, 110(6):068501, 2013.

\bibitem{bares2018_natcom}
J.~Bar{\'e}s, A.~Dubois, L.~Hattali, D.~Dalmas, and D.~Bonamy.
\newblock Aftershock sequences and seismic-like organization of acoustic events
  produced by a single propagating crack.
\newblock {\em Nature communications}, 9(1):1253, 2018.

\bibitem{Alava06_ap}
M.~J. Alava, P.~K. V.~V. Nukala, and S.~Zapperi.
\newblock Statistical models of fracture.
\newblock {\em Advances in Physics}, 55(3-4):349--476, 2006.

\bibitem{Bonamy08_prl}
D.~Bonamy, S.~Santucci, and L.~Ponson.
\newblock Crackling dynamics in material failure as the signature of a
  self-organized dynamic phase transition.
\newblock {\em Physical Review Letters}, 101(4):045501, Jul 2008.

\bibitem{bares14_prl}
Jonathan Bar{\'e}s, ML~Hattali, Davy Dalmas, and Daniel Bonamy.
\newblock Fluctuations of global energy release and crackling in nominally
  brittle heterogeneous fracture.
\newblock {\em Physical review letters}, 113(26):264301, 2014.

\bibitem{petri1994_prl}
A.~Petri, G.~Paparo, A.~Vespignani, A.~Alippi, and M.~Costantini.
\newblock Experimental evidence for critical dynamics in microfracturing
  processes.
\newblock {\em Physical Review Letters}, 73(25):3423, 1994.

\bibitem{ribeiro2015_prl}
H.~V. Ribeiro, L.~S. Costa, L.~G.~A. Alves, P.~A. Santoro, S.~Picoli, E.~K.
  Lenzi, and R.~S. Mendes.
\newblock Analogies between the cracking noise of ethanol-dampened charcoal and
  earthquakes.
\newblock {\em Physical review letters}, 115(2):025503, 2015.

\bibitem{brace1966stick}
W.~F. Brace and J.~D. Byerlee.
\newblock Stick-slip as a mechanism for earthquakes.
\newblock {\em Science}, 153(3739):990--992, 1966.

\bibitem{johnson2008effects}
P.~A. Johnson, H.~Savage, M.~Knuth, J.~Gomberg, and C.~Marone.
\newblock Effects of acoustic waves on stick--slip in granular media and
  implications for earthquakes.
\newblock {\em Nature}, 451(7174):57--60, 2008.

\bibitem{zapperi1997_nat}
S.~Zapperi, A.~Vespignani, and H.~E. Stanley.
\newblock Plasticity and avalanche behaviour in microfracturing phenomena.
\newblock {\em Nature}, 388(6643):658, 1997.

\bibitem{Papanikolaou2012_nat}
S.~Papanikolaou, D.~M. Dimiduk, W.~Choi, J.~P. Sethna, M.~D. Uchic, C.~F.
  Woodward, and S.~Zapperi.
\newblock Quasi-periodic events in crystal plasticity and the self-organized
  avalanche oscillator.
\newblock {\em Nature}, 490(517), 2012.

\bibitem{Urbach95_prl}
J.~S. Urbach, R.~C. Madison, and J.~T. Markert.
\newblock Interface depinning, self-organized criticality, and the barkhausen
  effect.
\newblock {\em Physical Review Letters}, 75:276--279, 1995.

\bibitem{Durin05_book}
G.~Durin and S.~Zapperi.
\newblock The barkhausen effect.
\newblock In G.~Bertotto and I.~Mayergoyz, editors, {\em The Science of
  Hysteresis}, page 181. Academic, New York, 2005.

\bibitem{ertacs1994_pre}
D.~Erta{\c{s}} and M.~Kardar.
\newblock Critical dynamics of contact line depinning.
\newblock {\em Physical Review E}, 49(4):R2532, 1994.

\bibitem{Planet09_prl}
R.~Planet, S.~Santucci, and J.~Ort\'in.
\newblock Avalanches and non-gaussian fluctuations of the global velocity of
  imbibition fronts.
\newblock {\em Phys. Rev. Lett.}, 102:094502, Mar 2009.

\bibitem{beggs2003_jn}
J.~M. Beggs and D.~Plenz.
\newblock Neuronal avalanches in neocortical circuits.
\newblock {\em Journal of neuroscience}, 23(35):11167--11177, 2003.

\bibitem{bellay2015_elife}
T.~Bellay, A.~Klaus, S.~Seshadri, and D.~Plenz.
\newblock Irregular spiking of pyramidal neurons organizes as scale-invariant
  neuronal avalanches in the awake state.
\newblock {\em Elife}, 4:e07224, 2015.

\bibitem{miller1996stress}
B.~Miller, Corey C.~O'Hern, and R.~P. Behringer.
\newblock Stress fluctuations for continuously sheared granular materials.
\newblock {\em Physical Review Letters}, 77(15):3110, 1996.

\bibitem{Hayman2011_pag}
N.~W. Hayman, L.~Duclou{\'e}, K.~L. Foco, and K.~E. Daniels.
\newblock Granular controls on periodicity of stick-slip events: Kinematics and
  force-chains in an experimental fault.
\newblock {\em Pure and Applied Geophysics}, 168(12):2239--2257, 2011.

\bibitem{pudasaini2007_book}
S.~P. Pudasaini and K.~Hutter.
\newblock {\em Avalanche dynamics: dynamics of rapid flows of dense granular
  avalanches}.
\newblock Springer Science \& Business Media, 2007.

\bibitem{lebouil_prl2014}
A.~Le Bouil, A.~Amon, S.~McNamara, and J.~Crassous.
\newblock Emergence of cooperativity in plasticity of soft glassy materials.
\newblock {\em Physical Review Letters}, 112:246001, Jun 2014.

\bibitem{bares2017_pre}
J.~Bar\'es, D.~Wang, D.~Wang, T.~Bertrand, C.~S. O'Hern, and R.~P. Behringer.
\newblock Local and global avalanches in a two-dimensional sheared granular
  medium.
\newblock {\em Physical Review E}, 96:052902, Nov 2017.

\bibitem{sethna2001crackling}
J.~P. Sethna, K.~A Dahmen, and C.~R. Myers.
\newblock Crackling noise.
\newblock {\em Nature}, 410(6825):242--250, 2001.

\bibitem{bi2011jamming}
D.~Bi, J.~Zhang, B.~Chakraborty, and R.~P. Behringer.
\newblock Jamming by shear.
\newblock {\em Nature}, 480(7377):355--358, 2011.

\bibitem{howell_prl1999}
D.~Howell, R.~P. Behringer, and C.~Veje.
\newblock Stress fluctuations in a 2d granular couette experiment: A continuous
  transition.
\newblock {\em Physical Review Letters}, 82:5241--5244, Jun 1999.

\bibitem{albert2001stick}
I.~Albert, P.~Tegzes, R.~Albert, J.~G. Sample, A.-L. Barab{\'a}si, T.~Vicsek,
  B.~Kahng, and P.~Schiffer.
\newblock Stick-slip fluctuations in granular drag.
\newblock {\em Physical Review E}, 64(3):031307, 2001.

\bibitem{petri_epjb2008}
A.~Petri, A.~Baldassarri, F.~Dalton, G.~Pontuale, L.~Pietronero, and
  S.~Zapperi.
\newblock Stochastic dynamics of a sheared granular medium.
\newblock {\em The European Physical Journal B}, 64(3-4):531--535, 2008.

\bibitem{lherminier2016granular}
S~Lherminier, R~Planet, G~Simon, M~M{\aa}l{\o}y, L~Vanel, and O~Ramos.
\newblock A granular experiment approach to earthquakes.
\newblock {\em Revista Cubana de F{\'\i}sica}, 33(1):55--58, 2016.

\bibitem{nasuno1998time}
S.~Nasuno, A.~Kudrolli, A.~Bak, and J.~P. Gollub.
\newblock Time-resolved studies of stick-slip friction in sheared granular
  layers.
\newblock {\em Physical Review E}, 58(2):2161, 1998.

\bibitem{nasuno1997friction}
S.~Nasuno, A.~Kudrolli, and J.~P. Gollub.
\newblock Friction in granular layers: Hysteresis and precursors.
\newblock {\em Physical Review Letters}, 79(5):949, 1997.

\bibitem{kaproth2013slow}
B.~M. Kaproth and C.~Marone.
\newblock Slow earthquakes, preseismic velocity changes, and the origin of slow
  frictional stick-slip.
\newblock {\em Science}, 341(6151):1229--1232, 2013.

\bibitem{krim2011stick}
J.~Krim, P.~Yu, and R.~P. Behringer.
\newblock Stick--slip and the transition to steady sliding in a 2d granular
  medium and a fixed particle lattice.
\newblock {\em Pure and applied geophysics}, 168(12):2259--2275, 2011.

\bibitem{dahmen2011simple}
K.~A.Dahmen, Y.~Ben-Zion, and J.~T. Uhl.
\newblock A simple analytic theory for the statistics of avalanches in sheared
  granular materials.
\newblock {\em Nature Physics}, 7(7):554--557, 2011.

\bibitem{aharonov2004stick}
E.~Aharonov and D.~Sparks.
\newblock Stick-slip motion in simulated granular layers.
\newblock {\em Journal of Geophysical Research: Solid Earth}, 109(B9), 2004.

\bibitem{liu2016driving}
C.~Liu, E.~E. Ferrero, F.~Puosi, J.-L. Barrat, and K.~Martens.
\newblock Driving rate dependence of avalanche statistics and shapes at the
  yielding transition.
\newblock {\em Physical review letters}, 116(6):065501, 2016.

\bibitem{white2003driving}
R.~A. White and K.~A. Dahmen.
\newblock Driving rate effects on crackling noise.
\newblock {\em Physical review letters}, 91(8):085702, 2003.

\bibitem{luan2004effect}
Binquan Luan and Mark~O Robbins.
\newblock Effect of inertia and elasticity on stick-slip motion.
\newblock {\em Physical review letters}, 93(3):036105, 2004.

\bibitem{lacombe2000dilatancy}
F.~Lacombe, S.~Zapperi, and H.~J. Herrmann.
\newblock Dilatancy and friction in sheared granular media.
\newblock {\em The European Physical Journal E: Soft Matter and Biological
  Physics}, 2(2):181--189, 2000.

\bibitem{zadeh2017avalanches}
A.~Abed Zadeh, J.~Bar{\'e}s, and R.~P. Behringer.
\newblock Avalanches in a granular stick-slip experiment: detection using
  wavelets.
\newblock {\em EPJ Web of Conference}, 140(03038), 2017.

\bibitem{Note1}
see SM for a full picture of the experiment (Fig. S1).

\bibitem{baldassarri2006brownian}
A.~Baldassarri, F.~Dalton, A.~Petri, S.Zapperi, G.~Pontuale, and L.~Pietronero.
\newblock Brownian forces in sheared granular matter.
\newblock {\em Physical review letters}, 96(11):118002, 2006.

\bibitem{Note2}
see SM for at at other $k$ values (Fig. S2).

\bibitem{pomeau1980intermittent}
Yves Pomeau and Paul Manneville.
\newblock Intermittent transition to turbulence in dissipative dynamical
  systems.
\newblock {\em Communications in Mathematical Physics}, 74(2):189--197, 1980.

\bibitem{strogatz2014nonlinear}
S.~H. Strogatz.
\newblock {\em Nonlinear dynamics and chaos: with applications to physics,
  biology, chemistry, and engineering}.
\newblock Westview press, 2014.

\bibitem{Note3}
see SM for unscaled curves (Fig. S3).

\bibitem{Note4}
see SM for $\protect \mathcal {P}_f$ and $P(f)$ for different $L$ (Fig. S5).

\bibitem{Note5}
see SM for non-scaled curves (Fig. S4).

\bibitem{Note6}
see SM for the calibration between $f(t)$ and $I(t)$ (Fig. S7).

\bibitem{Note7}
see SM for a comparison between $f(t)$ and $I(t)$ signals for different loading
  speeds (Fig. S6).

\bibitem{Note8}
see SM for $k=\infty $ PSD (Fig. S6).

\bibitem{bares2013_prl}
J.~Bar{\'e}s, L.~Barbier, and D.~Bonamy.
\newblock Crackling versus continuumlike dynamics in brittle failure.
\newblock {\em Physical review letters}, 111(5):054301, 2013.

\end{thebibliography}

\end{document}